\documentclass[11pt,letterpaper]{llncs}
\usepackage{amsmath}
\usepackage{amssymb}
\usepackage{graphicx}
\usepackage{marvosym}
\usepackage{url}
\usepackage{fancyhdr}

\usepackage{geometry}
\newcommand{\keywords}[1]{\par\addvspace\baselineskip
\noindent\keywordname\enspace\ignorespaces#1}

\fancypagestyle{firstpage}{\fancyhf{}
\setcounter{page}{33}
\fancyhead[C]{\small{International Journal of Network Security $ \& $ Its Applications (IJNSA) Vol.9, No.4, July 2017}}
\fancyfoot[L]{DOI: 10.5121/ijnsa.2017.9403}
\rfoot{\thepage}
}

\begin{document}


\title{\LARGE{Node Authentication Using BLS Signature in Distributed PKI Based MANETS}}


%
%
\author{\large{N Chaitanya Kumar \and Abdul Basit \and Priyadarshi Singh \and V. Ch. Venkaiah \and Y. V. Subba Rao}}
\institute{\large{School of Computer and Information Sciences, University of Hyderabad,\\ Hyderabad-500046, India}}

%


%
%


\maketitle

\thispagestyle{firstpage}

\begin{abstract}
Authenticating a node in mobile ad-hoc networks is a challenging task due to their dynamic and resource constraint infrastructure. For this purpose, MANETS adopt two kinds of approaches Public key cryptography and identity based cryptography. In Public Key Infrastructure (PKI), Certificate Authority (CA) is responsible for key management. In order to adopt it to MANET, the job of the CA must be distributed. The master secret key is shared among the nodes of the MANET, to self-organize the network without a central authority.The key is shared based on Shamir secret sharing scheme with bi-variate polynomial to make the MANET fully self-managed by nodes.In this paper, we considered PKI based scenario and proposed a new scheme to authenticate a node using BLS signature which is light weight compared to the existing schemes thus making it suitable for MANET.
\end{abstract}


\section{Introduction}

MANET known as Mobile Ad-Hoc Network is a self-organized, dynamic and infra-structureless network\cite{anjum}. MANET consists of mobile nodes that roam freely, every node has its own range of signal communication, other nodes within the range can interact and exchange messages. New nodes join and some other nodes may leave or some nodes fail to connect as they move out of the MANET network range\cite{daza}. The nodes in MANET are energy constrained, i.e., nodes are battery powered devices. There are many security threats to MANETS such as Denial of service, eavesdropping, interception and routing attacks\cite{perrig} \cite{perrig2}. Public Key Infrastructure (PKI)\cite{kent} helps in securing communication using authentication and encryption through digital certificates and public key cryptography respectively.The distributed PKI approach is adopted in this paper so as to make the MANET network completely de-centralized. 
\vspace{0.2cm} \\
Generally in a PKI environment, a certificate authority(CA) issues and manages the public key certificates of participating entities, the CA uses a master secret key $s$ to sign the certificate. General PKI is not suitable for MANET as we cannot assign the sole power of CA to a single node because of its dynamic and changing topology i.e., the node with CA functionality may break-down or move out of MANET range, which results in non-availability of CA. To achieve the distributed PKI environment for MANETS, we use a (t,n) threshold scheme\cite{zhou}\cite{blakley}\cite{shamir}, which helps in distributing CA power, i.e., we have to distribute the master secret key $s$ to nodes of the MANET\cite{moca}. In our proposal, we discuss how a threshold number of nodes sign a certificate and the verification of the certificate can be done by any node using BLS signature scheme\cite{bls}.

\subsection{Attacks on MANETS\cite{attack}}

In MANETS, there are two types of attacks- Passive and Active. Passive attacks capture valuable data in transit and active attacks cause huge damage to the network by disrupting the normal flow of the operations. Malicious nodes cause both active and passive attacks. A malicious node is the one, which does not authenticate itself to other honest nodes and misbehaves in the network. An honest node can also be compromised if it is under the control of the attacker. As the network comprises of layers of protocols, the attacks are specific to a layer and the security should also be implemented in the corresponding layer. Since the mobile nodes share a wireless medium, the messages transmitted can eavesdrop or fake messages may be injected at physical layer. Because of one-hop connectivity maintained among neighbors, the attacker can launch traffic analysis and traffic monitoring attacks. In network layer, the attacker exploits the routing algorithms to create routing hops and network congestion\cite{perrig2}. The attacker uses a compromised node to perform SYN flooding and denial of service(DOS) attacks at transport layer. The majority of attacks in the application layer are worm attacks, mobile viruses and repudiation attacks. Some attacks like denial of service and man-in-the-middle can be launched from several layers. This paper proposes node authentication using BLS signature, so that many of the attacks can be avoided.

\subsection{Distributed PKI}

Public key cryptography(PKC)\cite{william} provides many security services like confidentiality, integrity, authentication, non-repudiation, encryption and digital signatures. Public key infrastructure(PKI)\cite{kent} manages digital certificates which are important in the deployment of public key cryptography. In PKI environment, Certificate authority(CA) issues and maintains the certificates of participating entities, the certificate contains the public key and the ID of the entity, the CA signs the certificate using the master secret key $s$ and this certificate can be verified by the master public key $PK$. In MANETS we cannot adopt the same PKI, as the network is dynamic and infrastructure-less. So the role of the CA needs to be distributed to the nodes i.e., the master secret key $s$ is to be shared among different nodes and the master secret key can only be generated if atleast the threshold number of shares of secret are pooled together.

\subsection{Threshold Cryptography}

As MANET is a decentralized network, the master secret key $(s)$ of the PKI is distributed among the nodes using secret sharing schemes. One of the popular and most widely used secret sharing technique is the Shamir's secret sharing technique\cite{shamir}. In this scheme, dealer distributes a secret $s$ among n users. Each user receives it's share privately from the dealer. To reconstruct a secret, it uses (t, n) threshold access structure, where t out of n shares are required. Shamir's secret sharing scheme can be adopted in MANETS. Even the role of the dealer can be played by the nodes of MANET itself. This is achieved by using a bi-variate polynomial. This is discussed in section $3.1$.

\subsection{Related work}
One common issue faced by MANET when applying cryptography is, how to distribute the role of CA or trusted authority, many proposals use secret sharing technique to distribute secret key $s$ of CA or trusted authority to secure MANET. Zhou and Haas\cite{zhou} were the first to propose distributed CA for MANETS. They used threshold cryptography to distribute the role of the Certification Authority (CA) in a PKI scenario among a set of selected servers. However, this proposal is not suitable for a purely ad-hoc environment as these selected nodes may not always be available. Kong et al.\cite{kong} adapted a similar idea to distribute trust among all the nodes. However, their specific RSA threshold scheme has been proved insecure\cite{narasima}\cite{jarecki}. Shamir secret sharing technique\cite{shamir} is the most widely used secret sharing technique. We show that Shamir secret sharing technique along with the use of bi-variate polynomial helps to distribute the secret of CA among all nodes of MANET. In other works, bi-variate polynomials have already been used to dynamically allow new nodes joining the network without the need of any external trusted party. This technique is the result of inspiration from the original work of\cite{blundo}. Anzai et al.\cite{anzai} and Herranz et al.\cite{herranz} constructed decentralized, flexible, dynamic group key distribution schemes by using polynomials in two variables. The goal is to generate common group secret keys. Saxena et al.\cite{saxena} used similar technique to establish pairwise keys in a non-interactive way for a mobile ad-hoc scenario. Recently Daxing et al. \cite{Daxing} proposed aggregate signature algorithm for MANET using bilinear pairing and Hanaoka et al. \cite{Hanoka} construct multi user setting signature with tight security based on BLS signature.
\vspace{0.2cm} \\
Our work is more related to the cryptographic techniques proposed for MANETs by 
Herranz et al. \cite{herranz}. They proposed a fully self managed MANET and the ways to authenticate communication among the nodes. Our paper proposes the node authentication in their set up using BLS signature proposed by Boneh et al.\cite{bls}. Our proposal reduces the size of keys used as it uses the bilinear pairing. This scenario is much suitable for MANET because its nodes are mostly resource constraint devices and they can not afford the heavy computational overhead required by larger keys.

\section{preliminaries}

\subsection{Self-Organized PKI and Secret Sharing Technique}

In self-organized PKI for MANETS, the role of PKI is completely distributed among the nodes of MANET using secret sharing scheme\cite{shamir}. Blakley \cite{blakley} and Shamir \cite{shamir} were the first to introduce secret sharing techniques. In general a secret sharing scheme contains a dealer and a set \ $U= \{ {u_1,u_2,\cdots,u_n} \} $ of $n$ users. The dealer has a secret $S$ and wants to distribute the share $s_i$ of the  secret corresponding  to the  user $u_i$ privately. A valid subset  $u$ ( for : $u \subset U $) of atleast $t$ number of users holding valid shares can reconstruct the secret $S$. The $t$ is refereed as the threshold number and $(t,n)$ is refereed to as the threshold  access structure\cite{shamir}. In our paper, we use Shamir's secret sharing technique that uses a $(t,n)$ threshold access structure\cite{shamir}. Shamir's secret sharing scheme uses (t, n) threshold access structures using polynomial interpolation. Let $Z_q$ be a finite field with $q > n$ and let $S \in Z_q$ be the secret. The dealer picks a polynomial $P(x)$ of degree at most $t-1$, where the constant term of $P(x)$ is $S$ and all other coefficients are selected from $Z_q$ uniformly and independently at random. That is,
$$P(x)=S+ \sum_{\substack{i=1}}^{t-1}a_i*x^i$$

Every user $u_i$ is publicly associated to a field element $a_i$. Distinct parties are mapped to distinct field elements. The dealer privately sends  to user $u_i$ the value $[S]_i = P(a_i), for\ i = 1,2,\cdots, n$. Without loss of generality, we can assume that the set of parties willing to recover the secret $S$ is ${P_1,\cdots, P_t }$. The secret $S$ can obtained as $\sum_{\substack{i=1}}^{t}l_i*[s]_i$ where $l_i=\Pi_{j\neq i}\frac{a_j}{a_j-a_i}$ are the Lagrange coefficients. It is proven that any set of less than t parties obtain no information about $S$, that is, any secret is equally probable given their shares.

\subsection{Bilinear Pairing and Related Assumptions\cite{bilinear}}

Let $G_1$ be a cyclic additive group generated by some element P, whose order is a prime $q$, and $G_2$ be a cyclic multiplicative group of the
same order $q$. Let $a,b$ be elements of $Z_q^*$. We assume that
the discrete logarithm problem (DLP) in both $G_1$ and $G_2$ are
hard. A bilinear pairing is a map $e:G_1\times G_1\rightarrow
G_2$ with the following properties:

\begin{itemize}
\item Bilinear: For all S, T $\in G_1$ ,  $e(aS,bT)=e(S,T)^{ab}$.

\item Non-degenerate: There exists $S$ and $T\in G_1$ such that
$e(S,T)\neq 1$.

\item Computable: There is an efficient algorithm to compute $e(S,T)$
for all $S,T\in G_1$. 
\end{itemize}
We have the following assumptions:
\begin{itemize}

\item The Decisional Diffie-Hellman problem(DDHP) in $G_1$ should be easy.
\item The DDHP in $G_2$, the computational Diffie-Hellman problem(CDHP) and the
discrete logarithm problem (DLP) in both $G_1$ and $G_2$ should be hard.
\item The inversion of the bilinear pairing be hard,
i.e., the bilinear pairing inversion problem(BPIP) is
defined as:\par

\begin{itemize}
\item BPIP : Given $S\in G_1$ and $e(S,T)\in G_2$, find $T\in G_1$.
\end{itemize}
\end{itemize}

\subsection{BLS Signature\cite{bls}}
\label{sec:bls}
This scheme was introduced by D. Boneh, B. Lynn, H. Schacham. It is based on Computational Diffie-Hellman assumption on certain elliptic curve. We discuss the Gap Diffie-Hellman Group where this signature scheme works.
\subsubsection{Gap Diffie-Hellman Groups (GDH Groups)}
Consider a (multiplicative) cyclic group G = $\langle g\rangle$, with q = $\vert G\vert$ a prime. There are three problems on G.
\begin{itemize}
\item Group Action: Given $u, v$ $\in$ G, find $uv$.
\item Decision Diffie-Hellman : For a, b, c $\in Z^*_q$ , given $(g, g^a , g^b , g^c )$ decide whether c = ab.
\item Computational Diffie-Hellman : For a, b $\in Z^*_q$ , given $(g, g^a , g^b )$, compute $g^{ab}$.
\end{itemize}
The GDH group is defined as :

\begin{itemize}
\item  G is a $\tau$-decision group for Diffie-Hellman if the group action
can be computed in one time unit, and Decision Diffie-Hellman can be computed
on G in time at most $\tau$.
\item The advantage of an algorithm A in solving the Computational
Diffie-Hellman problem in a group G is
\\
$AdvCDH_A = Pr[A(g, g^a , g^b )] = g^{ab} : a, b \xleftarrow R $ $Z^*_q$
Where the probability is over the choice of a and b, and the coin tosses of A.
We say that an algorithm A ($t, \epsilon$)-breaks Computational Diffie-Hellman in G if
A runs in time at most $t$, and $AdvCDH_A\geq \epsilon $.
\item A prime order group G is a $(\tau, t, \epsilon )$-GDH group if it is a $\tau$-decision group for Diffie-Hellman and no algorithm $(\tau, \epsilon)$-breaks Computational
Diffie-Hellman on it.
\end{itemize}
\subsubsection{Signature Scheme}
\begin{itemize}
\item Setup of protocol:
\begin{description}
\item[Public information:] cryptographic hash function $H : \{0,1\}^* \to G_1$ and cryptographic bilinear map $e:G_1\times G_1 \to G_2$
\item[Signer's public key:]generator $P \in G_1, P_{pub} = sP$, where $s$ is the secret key and $P_{pub}$ is the public key.
\end{description}
\item Sign:For any message $M \in \{0,1\}^*$, signature is computed as
sig = sH(M)
\item Verify: Signature is only valid if the following equation holds.
$$e(P, sig) = e(P_{pub}, H(M)) $$
\item Proof:
$e(P, sig)=e(P, sH(M))=e(sP, H(m))=e(P_{pub}, H(m))$
\end{itemize}

\section{Our proposal}
This section is divided into four major phases namely Setup, Key Generation, Signature Generation Protocol and Signature Verification Protocol.

\subsection{Setup}

In this phase every node $n_i$ receives partial share $s_i$ of the MANET secret $s$. This is achieved using the following protocol.
\begin{itemize}
\item Let $n$ be the number of nodes in the MANET, $t$ be the threshold and $k$ be the founding number of nodes.
\item The founding number of nodes are such $t\leq k \leq n$.
\item Every founding node chooses a bi-variate polynomial $f_i(x,z)$, symmetric in $x,z$ and the max degree.
\item Every node $n_i$ computes $f_{ij}(h(n_j),z)$ for all other founding nodes and itself, $1\leq i\leq k.$
\item Now every node secretly sends computed $f_{ij}(h(n),z)$ to corresponding node $n_j$. Furthermore, node $n_i$ includes the value $y_i = f_i(0 )*P$ in each of these messages.
\item Finally every node has values received from other founding nodes and also it's own value $f_{ii}(h(n_i),z)$ with it. 
\\ Then every node $n_i$ computes  $f_i(z)=f(h(n_i),z)=\Sigma_{\substack{j \in k}}f_{ji}(h(n_i),z)$.
\item Now every node $n_i$ has partial secret $s_i=f_i(0)$ and a secret equation  $f(h(n_i),z)$.
\end{itemize}

The MANET secret function $f(x,z)=\Sigma_{i \in n}f_i(x,z)$ and MANET secret key is $s=f(0,0)$ are safe and hidden.This secret information can only be reconstructed if and only if there are at-least $t$ nodes having partial share of MANET secret. For a new node $n_w$ trying to join the network, it has to request at-least $t$ nodes for the values $f_{iw}(h(n_i),h(n_w))$. When $t$ nodes accept the node $n_w$ request, then they send $f_{iw}(h(n_i),h(n_w))$ to node $n_w$. Now node $n_w$ has $t$ values and these values are used in Lagrange's interpolation to derive a secret polynomial corresponding to node $n_w$, Lagrange's interpolation is applied as follows:
\begin{itemize}
\item  $$f_w(z)=f(h(n_w),z)=
\Sigma_{\substack{n_j \in n}}\Pi_{\substack{n_i \in n},n_i \neq n_j}\frac{(z-h(n_i))}{(h(n_j)-h(n_i))}*f(h(n_j),h(n_w))$$
\item The partial secret of node $n_w$ is $f_w(0)$ and secret polynomial of node $n_w$ is $f_w(z)\ i.e., f(h(n_w),z)$
\end{itemize}

\subsection{Key Generation}
After every node $n_i$ has received a partial secret $s_i$, now the nodes run RSA key generation protocol. The protocol is responsible for generating a public ($pk_i$) and private ($sk_i$) key pair. The private key ($sk_i$) is kept secret with the node $n_i$ and public key ($pk_i$) is made available to all other nodes. The public key $pk_i$ is used to encrypt messages that are sent to node $n_i$, and the node $n_i$ uses its private key $sk_i$ to decrypt messages as well as to sign messages.

%

\subsection{Signature Generation Protocol}
Now every node $n_i$ has two secret keys namely partial secret key of MANET $s_i$ and individual secret key $sk_i$, partial secret key is used to partially sign a certificate and any t out of n nodes are required to sign a certificate to generate fully signed/valid certificate. When a node $n_i$ wants to get a public key certificate, it asks its neighboring nodes to generate partial signature on the certificate linking $n_i$$||$$pk_i$. If the node $n_i$ receives at-least $(t-1)$ partial signs, then the node itself can generate a partial sign using it's own partial share, now the node has t partially signed values, then it uses the following Lagrange's interpolation to generate a fully signed certificate.

\begin{itemize}
\item $p_i = H(m)*s_i$ where $s_i$ is the individual share of each user and $H(m)$ is the hash of message m.
\item The final signature($shm$) is computed as 
 $ shm = \Sigma_{\substack{i \in t}} p_i*{L_i}$, where $L_i $ is Lagranges Coefficient.
 $L_i =\Pi_{\substack{p_j \in t},{j\neq i}}\frac{(0-h(N_j))}{(h(N_i)- h(N_j))}$
            
\end{itemize}
Now that every node obtains its certificate in the above described manner. Next we discuss the protocol to verify the certificate.

\subsection{Signature Verification Protocol}

Any node $n_j$ can verify the certificate of node $n_i$ by running the following protocol. Node $n_j$ has the following information regarding node $n_i$:
\begin{itemize}
\item the signed certificate of node $n_i$ ($shm$).
\item the public key of the MANET ($PK$) and value $P$.
\item ID of node $n_i$ and public key of node $n_i$ ($N_i$$||$$pk_i$).
\end{itemize}
The node $n_j$ uses BLS signature to verify the certificate:
\begin{itemize}    
   \item Verify $e(shm,P)=e(H(m),PK)$ If true certificate is valid, else invalid.    
\end{itemize}

\subsection{Example}
\begin{itemize}
	\item \textbf{Setup}
	\item Let the intial set of nodes $N_M = \{N_1,N_2,N_3,N_4\}$ \\ No.of Nodes = 4
	\item Public Parmeters : \\
        An additive group G of prime order q = 4019.\\
        - The curve used is     $E(F_{4019}) : y^2 = x^{3} +  1  $ \\
        - The Generator is P = E(3198,578)\\
 		- Let t = 2 (degree of polynomials) and k = 67 ( Field of Polynomials)
     \item An admissible bilinear pairing - Weil Pairing
     \item Two explicit collision resistant hash functions - HTP(Hash to Point) : 
     $\{0,1\}^*\rightarrow G_2$ and HTR(Hash to Range) : $\{0,1\}^*\rightarrow G_1$ where HTP 
     hashes the given message onto the elliptic curve group $G_2$ and HTR hashes the given value 
     to the group $G_1$.   
  
        \item Each node chooses a random symmetric-bivariate polynomial in GF(67)\\
        $N1 = 3x^2z+3z^2x+8xz+5z+5x+5$,
        $N2 = 5x^2z+5z^2x+3xz+8z+8x+9$\\
        $N3 = 8x^2z+8z^2x+5xz+3z+3x+6$,
        $N4 = 2x^2z+2z^2x+4xz+8z+8x+4$
        \item The implicit polynomial defined by all the nodes is \\
        F(x,z) = $ N_1 + N_2 + N_3 + N_4 $\\
               = $18x^2z + 18xz^2 + 20xz + 24x + 24z + 24$ 
    \item The secret $s$ of the MANET is F(0,0) = 24.
   \end{itemize}
       
\begin{itemize}
   \item Each node secretly sends to each of the other founding nodes the univariate polynomial 
   $F_{ij} = F_i(x,h(N_j)) $.
   \item The hash values of the nodes are \\
        $ h_{n1} = HTR('Node1',k) = 37 $, $ h_{n2} = HTR('Node2',k) = 54 $ \\
        $ h_{n3} = HTR('Node3',k) = 25 $, $ h_{n4} = HTR('Node4',k) = 17 $ 
     \end{itemize}
        
        \begin{itemize}
        \item Each node sends the following values to other Nodes :
        \item  N1 also includes $Y_1$ = 5 * P = (152,1437) \\
        $N_{11} = 44 x^{2} + 53 x + 56 $,
        $N_{12} = 28 x^{2} + 6 x + 7 $\\
        $N_{13} = 8 x^{2} + 3 x + 63 $,
        $N_{14} = 51 x^{2} + 3 x + 23 $
        \item N2 also includes $Y_2$ = 9 * P = (409,2266) \\
        $N_{21} = 51 x^{2} + 63 x + 37 $,
        $N_{22} = 2 x^{2} + 10 x + 39 $\\
        $N_{23} = 58 x^{2} + 59 x + 8 $,
        $N_{24} = 18 x^{2} + 30 x + 11 $    
        \item N3 also includes $Y_3$ = 6 * P = (3063,3143) \\
        $N_{31} = 28 x^{2} + 18 x + 50 $,
        $N_{32} = 30 x^{2} + 17 x + 34 $\\
        $N_{33} = - x^{2} + 36 x + 14 $,
        $N_{34} =  2 x^{2} + 55 x + 57 $
        \item N4 also includes $Y_4$ = 4 * P = (3863,2497) \\
        $N_{41} = 7 x^{2} + 13 x + 32 $,
        $N_{42} = 41 x^{2} + 26 x + 34 $\\
        $N_{43} = 50 x^{2} + 18 x + 3 $,
        $N_{44} = 34 x^{2} + 51 x + 6$            
        \item Then all the nodes calculate their secret univariate polynomial from the recieved values.
        \item $S_1(x) =63 x^{2} + 13 x + 41 $ ,
         $S_2(x) = 34 x^{2} + 59 x + 47 $
        \item $S_3(x) = 48 x^{2} + 49 x + 21 $,
         $S_4(x) = 38 x^{2} + 5 x + 30 $ 
        
        \end {itemize}
        
        \begin{itemize}
        \item The public key, PK = s * P \\  $  \hspace{3cm} $   = 24 * E(3198,578) = E(2651, 2267)
        \item PK should also equal to $Y_1 + Y_2 + Y_3 + Y_4$\\
        =E(152,1437)+E(409,2266)+E(3063,3143)+E(3863,2497) = E(2651, 2267)
        \item Each node calculates its share from $S_i(0)$.\\
        The shares of the nodes are - $S_1 = 41, S_2 = 47 ,S_3 = 21, S_4 = 30$
        \item These shares can be verified by substituting hash value of the nodes in the following polynomial f(z) = F(0,z)  \hspace{0.6cm} = $ 24*z + 24$
        \end{itemize}

        \begin{itemize}
        \item If Node $N_5$ wants to join the MANET, It should identify it self to 3 other nodes and request for acceptance. $\{N_2,N_3,N_4\}$ \\
        $h_{n5}= HTR('Node5',k) = 27$
        \item $N_5$ receives the following values \\
        $ S_{25} = S_2(27)\ mod\  67= 28  $,
        $ S_{35} = S_3(27)\ mod\  67=22 $\\
        $ S_{45} = S_4(27)\ mod\  67=62  $
        \item $N_5$ computes its secret univariate polynomial by using Lagrange interpolation 
        $S_5(x) =  17*x^2 + 18*x + 2 $
        \end{itemize}

\begin{itemize}
\item \textbf{Key Generation}
            \item Each node computes its own key pair as follows :\\
            Node 1 = [(89,649),(189,649)],
            Node 2 = [(17,321),(25,321)]\\
            Node 3 = [(63,115),(7,115)],
            Node 4 = [(91,202),(11,202)]
            
    \item \textbf{Signature Generation}
                \item The share of each node in MANETs secret key is used as secret key for signature i.e 
            $S_1 = 41, S_2 = 47 ,S_3 = 21, S_4 = 30$
            \item Each node produces a certificate by linking Id with PK\\
            $m_1$='Node1'+'89'+'649',
            $m_2$='Node2'+'17'+'321'\\
            $m_3$='Node3'+'63'+'115',
            $m_4$='Node4'+'91'+'202'
            \item Then all the nodes exchange partial signatures to compute fully signed certificate.
            \end{itemize}
    \begin{itemize}
    \item If Node 1 wants to compute its certificate ($m_1$='Node1'+'89'+'649'), then it requests Node 2, Node 3 and Node 4 for their partial signatures.
    
            \item Here P = E(3198,578), s = 24, mpub =E(2651,2267) and s2 = 47,s3 = 21,s4 = 30
            \item $hm_1 =  HTP(m_1) = E(163,1362)$
            \item The partial signatures of Nodes 2, 3 and 4 are \\
             $p_2=(hm_1)*{s2} $,
             $p_3=(hm_1)*{s3}  $,
$p_4=(hm_1)*{s4}  $
            \item By lagranges interpolation we get the signature on the message1 as shm1=E(2350,3239).
            \item \textbf{Signature Verification}
            \item Calculate  e(hm1,mpub)=1365*a + 2045
            \item Calculate e(shm1,P)=1365*a + 2045 this is equal to e(hm1,mpub)
            \item Hence Verified
            \item Message communication after verification

            \item Public and private key pairs of each node\\
            Node 1 = [(e1,n1),(d1,n1)]=[(89,649),(189,649)]\\
            Node 2 = [(e2,n2),(d2,n2)]=[(17,321),(25,321)]\\
            Node 3 = [(e3,n3),(d3,n3)]=[(63,115),(7,115)]\\
            Node 4 = [(e4,n4),(d4,n4)]=[(91,202),(11,202)]\\
            Message (M)= 56
        \item If Node 1 wants to send a message to Node 3,then Node 1 Encrypts the message using Node 3's public key and sends to Node 3.
            \item C=Encrypt(M,e3,n3) 
            $C= (mod(56,115)^{63})$\\
            Encrypted Value C = 463
            \item Node 3 receives the Cipher value and Decrypts the message using Node 3 private key.
            \item M=Decrypt(C,d3,n3) $m=mod(463,115)^7$\\
            Decrypted Value M=56

        \end{itemize}      
        
%

\section{Conclusion}
In this paper, we proposed a new scheme of verifying a certificate in decentralized PKI based MANETS. In our scheme the nodes of the MANET holds a secret share and every node chooses its own public and private keys. The public key is associated with the node identity in the certificate. This certificate management is done using BLS Signature. Our scheme uses a bivariate polynomial to reduce the communication overhead. The same technique can be used in performing other functionalities of MANET like implementing threshold operations in sub group nodes communication and share verification etc.

\vspace{2cm}

\section*{Authors}
\noindent {\bf N Chaitanya Kumar} received M.Tech from JNTU Hyderabad, and he did Bachelor degree in computer science. Currently, he is pursuing his PhD in Computer Science
from the University of Hyderabad. His research interests include Information security, Cryptography in MANET.\\

\noindent {\bf Abdul Basit} received Master of computer application from Jamia Hamdard University New Delhi. He did Bachelor of Science in Information technology from SMU Gangtok. Currently, he is pursuing his PhD in Computer Science from the University of Hyderabad. His research interests include Information security, Cryptography, Cyber security, and Algorithms.\\

\noindent {\bf Priyadarshi Singh} received M.Tech from IIT(ISM) Dhanbad. He did Bachelor degree in Information Technology. Currently, he is pursuing his PhD in Computer Science from the University of Hyderabad. His research interests include Cryptography, Public key infrastructure.\\

\noindent {\bf V. Ch. Venkaiah} obtained his PhD in 1988 from the Indian Institute of Science (IISc), Bangalore in the area of scientific computing. He worked for several
organisations including the Central Research Laboratory of Bharat Electronics, Tata Elxsi India Pvt. Ltd., Motorola India Electronics Limited, all in Bangalore.
He then moved onto academics and served IIT, Delhi, IIIT, Hyderabad, and C R Rao Advanced Institute of Mathematics, Statistics, and Computer Science. He is currently serving the Hyderabad Central University. He is a vivid researcher. He designed algorithms for linear
programming, subspace rotation and direction of arrival estimation, graph colouring, matrix symmetriser, integer factorisation, cryptography, knapsack problem, etc.\\

\noindent {\bf Subba Rao Y V} obtained his PhD from the University of Hyderabad. Currently, he is an Assistant Professor in the School of Computer and Information Sciences,
University of Hyderabad. His area of interests includes Cryptography, Theory of Computation, etc.

\end{document}